\documentclass{article}
\usepackage{graphicx} 
\usepackage[a4paper, total={4in, 8in}]{geometry}
\usepackage{url}
\usepackage{hyperref}
\usepackage[
sorting=none
]{biblatex}
\usepackage{listings}
 \title{End to end Hindi to English speech conversion using Bark, mBART and a finetuned XLSR Wav2Vec2 \\\line(1,0){\textwidth}}
\begin{document}
\author{
  Aniket Tathe\\
    \texttt{\scriptsize Department of  Mechanical Engineering}\\
  \texttt{\scriptsize MES College of Engineering, Pune, India}\\
   \texttt{\scriptsize  anikettathe.08@gmail.com}
  \and
  Anand Kamble\thanks{Corresponding author. Email: \scriptsize amk23j@fsu.edu}\\
  \texttt{\scriptsize Department of Scientific Computing}\\
  \texttt{\scriptsize Florida State University, USA}\\
  \texttt{\scriptsize amk23j@fsu.edu}
  \and
  Suyash Kumbharkar\\
    \texttt{\scriptsize Department of Electrical Engineering }\\
    \texttt{\scriptsize and Information Technology}\\
  \texttt{\scriptsize Technische Hochschule Ingolstad, Germany}\\
  \texttt{\scriptsize suk9387@thi.de}
  \and
  \\
  Atharva Bhandare\\
    \texttt{\scriptsize Department of  Mechanical Engineering}\\
  \texttt{\scriptsize MES College of Engineering, Pune, India}\\
  \texttt{\scriptsize atharvabhandare512@gmail.com}
  \and
  \\
  Anirban C. Mitra\\
    \texttt{\scriptsize Department of  Mechanical Engineering}\\
  \texttt{\scriptsize MES College of Engineering, Pune, India}\\
  \texttt{\scriptsize amitra@mescoepune.org}
}

\date{}

\maketitle
\section{Abstract}
Speech has long been a barrier to effective communication and connection, persisting as a challenge in our increasingly interconnected world. This research paper introduces a transformative solution to this persistent obstacle – an end-to-end speech conversion framework tailored for Hindi-to-English translation, culminating in the synthesis of English audio. By integrating cutting-edge technologies such as XLSR Wav2Vec2 for automatic speech recognition (ASR), mBART for neural machine translation (NMT), and a Text-to-Speech (TTS) synthesis component, this framework offers a unified and seamless approach to cross-lingual communication. We delve into the intricate details of each component, elucidating their individual contributions and exploring the synergies that enable a fluid transition from spoken Hindi to synthesized English audio.

\section{Keywords}
XLSR Wav2Vec2, mBART, Bark, End-to-end Speech Conversion, Common Voice Corpus.

\section{Introduction}
In the intricate tapestry of global communication, speech has persisted as a formidable barrier, hindering effective connection and understanding in our increasingly interconnected world. Despite ongoing research efforts, addressing this challenge becomes more intricate when dealing with low-resource Automatic Speech Recognition (ASR) languages. A variety of open-source ASR models, including XLSR Wav2Vec2\cite{babu22_interspeech}, Whisper\cite{radford2022robust} and Kaldi\cite{kaldi}, are available, each with its own strengths and weaknesses based on factors such as usability, speed, accuracy, and the intended task. Similarly, the landscape of Neural Machine Translation (NMT) introduces diverse options like mBart\cite{10.1162/tacl_a_00343}, MarianMT\cite{marianMT} and T5\cite{48643}, each offering unique advantages. Furthermore, Text-to-Speech (TTS) models play a crucial role in speech-enabled applications that require converting text to speech, simulating the nuances of the human voice. Models such as Tortoise TTS\cite{Betker_TorToiSe_text-to-speech_2022}, Bark\cite{suno-bark}, Tachotron\cite{wang2017tacotron}, Tachotron 2\cite{8461368} and FastSpeech\cite{NEURIPS2019_f63f65b5},FastSpeech2\cite{ren2022fastspeech} contribute to this realm. This research endeavors to address the persistent challenge of cross-lingual communication by proposing a novel approach to convert spoken Hindi into understandable English. Our devised system integrates cutting-edge technologies, employing a finetuned XLSR Wav2Vec2 for speech recognition, mBART for language translation, and Bark for audio handling. In the following sections, we explore the intricacies of each model, shedding light on their distinctive contributions and demonstrating how, together, they transform spoken Hindi into coherent English audio. 

\section{Methodology}
\subsection{XLSR Wav2vec2 Fine-tuning}
The Common Voice corpus\cite{ardila2020common} stands as an extensive and remarkably diverse repository of transcribed speech, featuring an impressive collection of 19,160 validated hours across 114 languages. Each dataset entry comprises a distinctive MP3 audio file along with its corresponding text transcript. Additionally, the resource includes valuable demographic metadata, encompassing details such as age, gender, and accent, in a substantial portion of the 28,751 recorded hours. This supplementary metadata proves invaluable for augmenting the accuracy of speech recognition systems. Despite its vast size and inclusivity, the Common Voice corpus poses challenges when dealing with low-resource automatic speech recognition (ASR) languages like Hindi. In the most recent release, Common Voice 16.1, English language data spans approximately 3,438 hours of audio, with 2,586 hours validated by a community of 90,474 contributors.
In contrast, the Hindi language section comprises only about 21 hours of recorded audio, of which 14 hours have undergone meticulous validation. The Common Voice 13 Hindi dataset was utilized for the fine-tuning of the XLSR Wav2Vec2 model. This subset contained 19 hours of recorded audio, with 14 hours thoroughly validated.

The XLSR Wav2Vec2 model, an acronym for "Cross-lingual Self-supervised Representations from Wav2Vec2" is a pivotal component in our speech recognition methodology. Developed by Facebook AI Research (FAIR), this robust model has undergone extensive training on a multilingual dataset, boasting an impressive training duration of 53,000 hours of unlabeled speech. Designed to support a wide spectrum of languages, the model was trained on a diverse set of 53 languages, emphasizing its cross-lingual capabilities. Its proficiency in understanding the intricacies of spoken languages, including Hindi, makes it a cornerstone in our end-to-end speech conversion framework. With its extensive training duration and multilingual support, it significantly contributes to the system's ability to comprehend diverse spoken languages, thereby enhancing the overall effectiveness of our speech conversion framework. 

The model that was used for fine-tuning was "\href{https://huggingface.co/facebook/wav2vec2-large-xlsr-53}{facebook/wav2vec2-large-xlsr-53}". The model was trained on NVIDIA A5000 GPU and it was trained for 60 epochs with a learning rate of $1 \times 10^{-4}$ and a weight decay of $2.5 \times 10^{-6}$ after testing with numerous values. Various values like $3 \times 10^{-4}$, $2 \times 10^{-6}$, $5 \times 10^{-5}$, $1 \times 10^{-6}$, etc., were tested for the learning rate. The model performed a little better with weight decay, though the difference wasn't significant. The model gave better accuracy and WER (0.428) after training for 60 epochs continuously rather than training for 30 epochs with $1 \times 10^{-6}$ (learning rate) and zero weight decay and the rest 30 epochs for $1 \times 10^{-8}$ (learning rate) and $2.5 \times 10^{-6}$ (weight decay). The WER graph can be seen below in figure \ref{fig:enter-label}. This trained model can be found at "\href{https://huggingface.co/Aniket-Tathe-08/XLSR-Wav2Vec2-Finetuned}{Aniket-Tathe-08/XLSR-Wav2Vec2-Finetuned}" which accepts Hindi audio as input and outputs corresponding Hindi text.

\begin{figure*}[ht]
    \centering
    \includegraphics[width=0.5\textwidth]{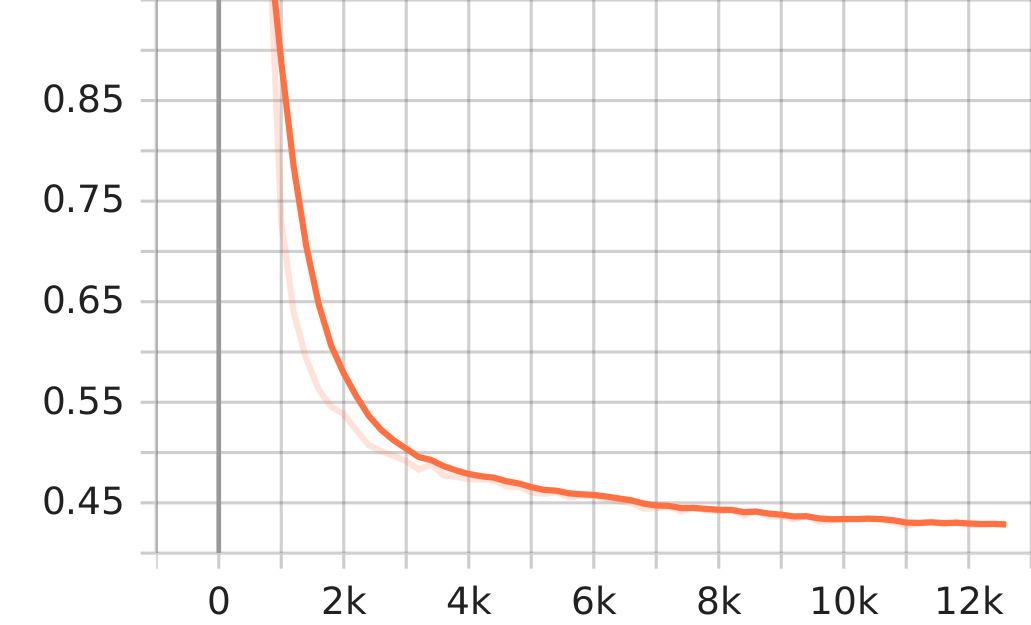} 
    \caption{WER (Word error rate)}
    \label{fig:enter-label}
\end{figure*}

\subsection{Neural Machine Translation using mBART}

At the heart of our neural machine translation (NMT) approach is mBART—short for Multilingual BART—meticulously developed by Facebook AI Research (FAIR). Acting as an extension of the BART (Bidirectional and Auto-Regressive Transformers) model, mBART is tailored with precision for the complex task of multilingual translation. Its standout feature lies in its remarkable ability to handle multiple languages within one unified model, making it an efficient and versatile solution for our cross-lingual speech translation system. In our research framework, mBART plays a central role as the neural machine translation component, seamlessly transforming recognized Hindi text into English.

\subsection{Bark}

Bark, developed by Suno, is an innovative text-to-audio model that utilizes transformer-based technology to produce natural-sounding speech in multiple languages. This advanced system is not limited to speech generation but can also create various types of audio content, including music, ambient sounds, and basic sound effects. Additionally, Bark has the ability to generate nonverbal expressions such as laughter, sighs, and crying, adding a new dimension to its audio capabilities. The model is accessible through the Hugging Face Space by Suno and can be applied in diverse fields such as podcast creation, audiobook production, and the development of sound effects for various applications. By leveraging GPT-style models, Bark can generate highly expressive and emotive voices with minimal adjustments, capturing subtle nuances such as tone, pitch, and rhythm. Furthermore, it supports multiple languages, demonstrating exceptional clarity and accuracy in speech generation across different linguistic contexts. Suno also provides access to pre-trained model checkpoints to facilitate research and development. However, it is important to be mindful of the potential dual use of such technology, and Suno has taken steps to mitigate unintended use by offering a classifier that can accurately detect Bark-generated audio.
In the process of employing Bark, we seamlessly convert English text to audio. Bark offers 10 different prompts (voices) for the English language that can be specified using the “history\_prompt” argument during input. For those seeking custom voice generation, the Serp-ai\cite{Serp-ai} repository can be utilized. Although alternatives like Tortoise-TTS, Tacotron, and similar TTS systems are available, Bark's selection is attributed to its multilingual capabilities, accuracy, and the ability to even voice clone a custom voice.
\\
The flowchart of the entire end-to-end speech translation system can be viewed below in figure \ref{fig:flowcart}.
\begin{figure}[h]
    \centering
    \includegraphics[width=\textwidth]{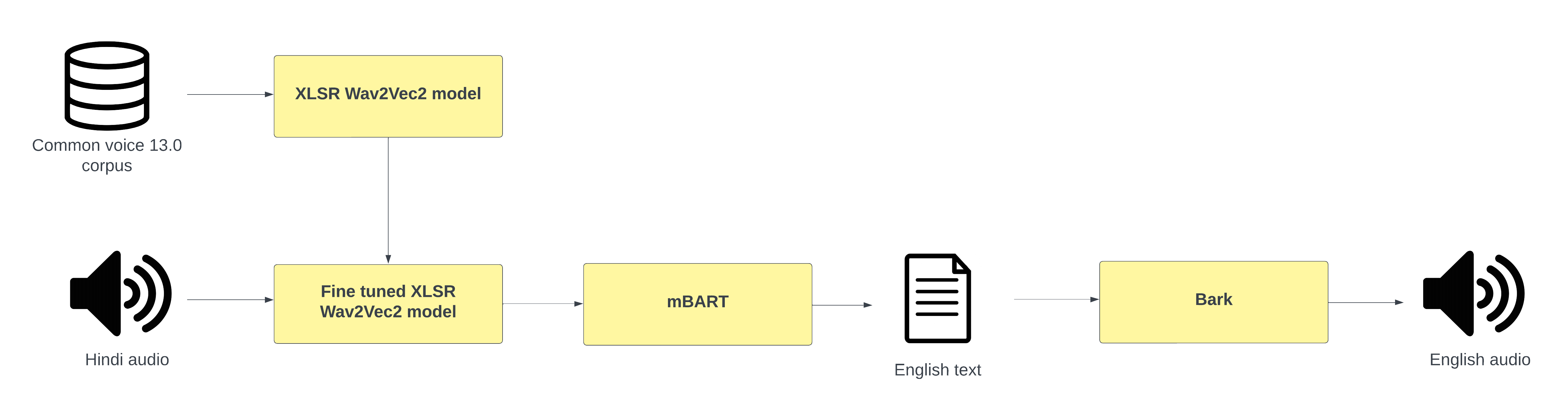}
    \caption{Hindi to English speech conversion}
    \label{fig:flowcart}
\end{figure}

\subsection{Results and Discussion}
The culmination of our research presents a robust end-to-end speech conversion framework for translating audio from Hindi to English, employing Bark, mBART, and a fine-tuned XLSR Wav2Vec2. Throughout our exploration, the seamless collaboration of these components provided a unified solution, addressing the challenges of cross-lingual speech conversion. The research outcomes not only advance the field of speech-to-text and machine translation but also offer practical applications in diverse domains such as podcast creation, audiobook production, and sound effects development.  Furthermore, the potential of such a system extends beyond research applications. This system can be utilized to create portable speech translation devices by incorporating additional components such as a microphone, a speaker, and a Raspberry Pi. This could facilitate communication for many people, including tourists. It is worth noting, however, that the implementation of such a device might pose challenges due to the computational intensity of the models involved. While there are smaller and faster models available, attempting to create a compact and efficient device remains a task that demands careful consideration and optimization. Despite these challenges, the exploration of smaller and faster models signifies a promising avenue for future endeavors in developing accessible and practical speech translation solutions.

\nocite{9906230}
\nocite{yi2021applying}
\nocite{10095644}
\nocite{shahgir2022applying}
\nocite{peng21e_interspeech}
\nocite{10095036}
\nocite{9915932}
\nocite{smith2018disciplined}

\printbibliography[title={References}]

\end{document}